# Double-Pulse Generation of Indistinguishable Single Photons with Optically Controlled Polarization

Junyong Yan[1,‡], Shunfa Liu[2,‡], Xing Lin[1], Yongzheng Ye[1,3], Jiawang Yu[1,4], Lingfang Wang[1], Ying Yu[2,*], Yanhui Zhao[2,5], Yun Meng[6,7], Xiaolong Hu[6,7], Dawei Wang[1,8], Chaoyuan Jin[1,4], Feng Liu[1,4,†]

[1]Interdisciplinary Center for Quantum Information, State Key Laboratory of Modern Optical Instrumentation, College of Information Science and Electronic Engineering, Zhejiang University, Hangzhou 310027, China.

[2]State Key Laboratory of Optoelectronic Materials and Technologies, School of Physics, School of Electronics and Information Technology, Sun Yat-sen University, Guangzhou 510275, China.

[3]College of Optical Science and Engineering, Zhejiang University, Hangzhou 310027, China.

[4]International Joint Innovation Center, Zhejiang University, Haining 314400, China.

[5]School of Physics and Optoelectronic Engineering, Ludong University, Yantai, Shandong 264025, China.

[6]School of Precision Instrument and Optoelectronic Engineering, Tianjin University, Tianjin 300072, China.

[7]Key Laboratory of Optoelectronic Information Science and Technology, Ministry of Education, Tianjin 300072, China.

[8]Zhejiang Province Key Laboratory of Quantum Technology and Device, Department of Physics, Zhejiang University, Hangzhou 310027, China.

†feng_liu@zju.edu.cn; *yuying26@mail.sysu.edu.cn

**ABSTRACT:** Single-photon sources play a key role in photonic quantum technologies. Semiconductor quantum dots can emit indistinguishable single photons under resonant excitation. However, the resonance fluorescence technique typically requires cross-polarization filtering which causes a loss of the unpolarized quantum dot emission by 50%. To solve this problem, we demonstrate a method to generate indistinguishable single photons with optically controlled polarization by two laser pulses off-resonant with neutral exciton states. This scheme is realized by exciting the quantum dot to the biexciton state and subsequently driving the quantum dot to an exciton eigenstate. Combining with magnetic field, we demonstrated the generation of photons with optically controlled polarization (polarization degree of 101(2)%, laser-neutral exciton detuning up to 0.81 meV, high single-photon purity (99.6(1)%) and indistinguishability (85(4)%). Laser pulses can be blocked using polarization and spectral filtering. Our work makes an important step towards indistinguishable single-photon sources with near-unity collection efficiency.

In recent years, photonic quantum technologies have attracted much attention due to their great potential in quantum computing/simulation[1–4], quantum communication[5–8], and quantum metrology/sensing[9–12]. Devices that can deterministically generate photons with high single-photon purity, indistinguishability, and photon collection efficiency are crucial for scalable photonic quantum technologies. Semiconductor quantum dots (QDs) are one of the most promising candidates for on-demand single-photon sources (SPSs). After decades of developments, near-unity single-photon purity (99.1%[13]) and indistinguishability (99.5%[14]) have been achieved with coherently driven single QDs in optical cavities[15–18] or waveguides[19,20]. However, despite the nearly-ideal single photons, to data, the highest collection efficiency of polarized single photons coupled into a single-mode fiber is limited to 57%[16]. Further improvement is still needed to meet the requirements for scalable photonic quantum technologies.

One of the limiting factors of the collection efficiency for QD SPSs is the excitation scheme. The indistinguishability of photons emitted from a QD is negatively affected by the uncertainty of the photon emission time[21]. Therefore, a QD SPS is typically operated with a resonance fluorescence technique where the QD is directly excited to the lowest exciton states to avoid any time jitter. The weakness of this method is that to get rid of the resonant laser, cross-polarization filtering, which blocks at least 50% of the unpolarized QD emission, has to be used. Various excitation schemes have been explored to solve this problem, for example, by using a pair of dichromatic pulses 0.27 meV detuned from the QD emission[22,23], longitudinal acoustic phonon-assisted excitation with exciton population up to 85%[24–26], optical microcavities with nondegenerate linearly polarized modes[16,27] and stimulated two-photon emission from the biexciton state via a virtual state[28,29]. However, an excitation scheme that deterministically generates indistinguishable single photons with a near-unity polarization degree and large laser-QD detuning is still missing. Combining coherent two-photon excitation (TPE) [30–32] with stimulated emission of the biexciton via neutral exciton states [33,34] (see Fig. 1(a)) could potentially meet all the requirements above, but this method has not been employed to generate indistinguishable single photons.

In this work, we demonstrate the on-demand generation of indistinguishable single photons with optically controlled polarization using two laser pulses spectrally separated from the neutral exciton ($X$) state. The main advantage of this scheme is that the excitation laser can be blocked by combining spectral and polarization filtering or, in principle, only using an ultra-narrow-bandwidth spectral filter. With the help of the magnetic field, the neutral exciton transition of the In(Ga)As QD was tuned in resonance with the micropillar cavity, and the laser-neutral exciton detuning was increased to 0.81 meV. Under this optimal condition, single photons were generated with a degree of polarization (DOP) up to 101(2)%, single-photon purity of 99.6(1)%, and indistinguishability of 85(4)%. The high photon polarization degree and relatively large laser-neutral exciton detuning make it more practical to realize a high-performance QD SPS without scarifying the photon collection efficiency.

The physical principle of the double-pulse excitation (DPE) scheme is depicted in Fig. 1(a). The QD is coherently excited from the ground ($G$) state to the biexciton ($XX$) state by a two-photon excitation (TPE) $\pi$ pulse (see the orange arrows). Then the QD is triggered to one of the linearly polarized neutral exciton eigenstates, e.g. $X_H$ ($X_V$), by the second $\pi$ pulse resonant with the $X$-$XX$ transition (see the gray arrow). Finally, the QD emits a horizontally (vertically) polarized photon via spontaneous emission. The trigger pulse is critical for realizing a high-performance SPS. Firstly, it eliminates the time jitter caused by the relaxation from the $XX$ to $X$ states [21]. This, in theory, ensures that the indistinguishability can reach a value as high as that obtained with the conventional resonance fluorescence technique [35]. Secondly, the QD is deterministically driven to an exciton eigenstate, preventing the emission of photons with different polarizations. Thanks to the biexciton binding energy $E_b$, the TPE and trigger pulses are spectrally detuned from the $X$ transition by $E_b/2$ and $E_b$, respectively (see Fig. 1(b)). This detuning allows laser pulses to be blocked using an ultranarrow bandwidth spectral filter. Furthermore, the TPE pulse which is closer to the $X$ transition can also be filtered out by a polarizer parallel to the $X_H$ ($X_V$) states, relaxing the requirement for the spectral filter.

The DPE scheme was demonstrated using a QD in a micropillar cavity (see the inset in Fig. 1(c)). The reflection spectrum of the cavity and the photoluminescence (PL) spectrum of the In(Ga)As QD under the two-photon excitation are shown in Fig. 1(c). The biexciton binding energy $E_b$, corresponding to the energy difference between the $X$ and $XX$ peaks, was 0.90 meV. The exciton fine-structure splitting (FSS) was 5.9(1) μeV (see Supporting Fig. 1), much smaller than the linewidth of the trigger pulse (0.12 meV). Compared with the $XX$ peak, the $X$ peak was more enhanced due to a smaller detuning relative to the cavity resonance. The micropillar cavity (Q factor: ~1351) has two degenerate fundamental modes which are slightly elliptically polarized (see Supporting Fig. 2).

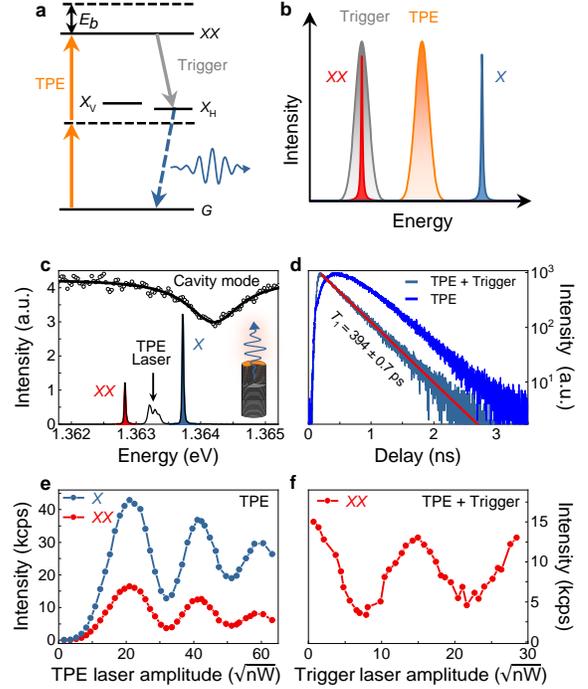

Figure 1. **a** Energy-level scheme of a QD. The QD is deterministically excited to the biexciton ($XX$) state from the ground ($G$) state by the two-photon excitation (TPE) pulse (orange arrow) and subsequently driven to one of the neutral exciton eigenstates by the trigger pulse (gray arrow). Finally, the QD emits a polarized photon via spontaneous emission (blue dashed arrow). $X_H$ / $X_V$: the horizontally/vertically polarized neutral exciton states, respectively. $X_H$ and $X_V$ are split by the fine-structure splitting (FSS). **b** The illustration of the DPE scheme in the form of a photoluminescence (PL) spectrum. Red/blue peak: biexciton/exciton emission. Orange/gray peak: TPE/trigger pulse. **c** PL spectrum of the QD under TPE. Red/blue peak: biexciton/exciton emission. Black open circles: reflection spectrum of the cavity measured with a halogen lamp. Black line: fitting with a Lorentzian function. Inset: schematic of a QD in micropillar cavity. **d** PL decays of the $X$ emission under TPE (blue) and DPE (light blue). The $X$ lifetime $T_1$ measured with DPE was 394(1) ps, ~26 times longer than the trigger pulse duration (15 ps). Red line: fitting with a single exponential function. **e** Intensities of the $X$ emission (blue circles) and $XX$ emission (red circles) as a function of the electric field amplitude of the TPE pulse, showing a clear Rabi oscillation. **f** The intensity of the $XX$ emission as a function of the electric field amplitude of the trigger pulse.

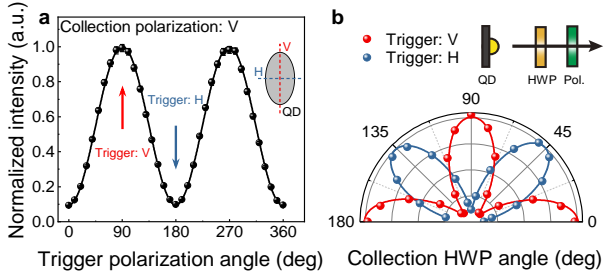

**Figure 2. a** The intensity of the vertically polarized component of the *X* emission as a function of the trigger pulse polarization angle. Inset: quantization axis of the quantum dot. **b** The intensity of the *X* emission as a function of half-wave plate (HWP) angle measured with vertically (red) polarized trigger pulse and horizontally (light blue) polarized trigger pulse. Inset: experimental setup for characterizing the polarization of the QD emission. Pol.: polarizer.

Before we move to the measurement of the indistinguishability, a key step is to show that the trigger pulse can eliminate the time jitter induced by spontaneous emission from the *XX* to *X* states. To this end, the exciton lifetime was measured with and without the trigger pulse (see Fig. 1(d)). With only a TPE pulse, the PL decay curve (blue line) showed a clear long rising edge (250 ps defined by 1/e of the PL maximum) corresponding to the relaxation from the *XX* to *X* states. By applying the trigger pulse, the long rising edge was transformed to a sharp edge (62 ps, limited by the instrument response time of 60 ps) proving the time jitter was indeed efficiently eliminated by the trigger pulse.

In addition to avoiding the time jitter, another important criterion of an excitation scheme for a deterministic SPS is the excitation efficiency $\eta$. $\eta$ is defined by the probability of obtaining one photon in an excitation cycle. To obtain a high excitation efficiency, the two pulses in the DPE scheme need to be $\pi$ pulses. The pulse area of the TPE and trigger pulses were characterized by performing Rabi oscillation measurements. Fig. 1(e) shows the intensity of the *X* and *XX* emission as a function of the TPE pulse amplitude.

The Rabi oscillation of the *G-XX* transition was clearly observed. The TPE pulse at the first maximum of the PL intensity corresponds to a $\pi$ pulse. Next, we excited the QD with a TPE $\pi$ pulse and monitored the intensity of the *XX* emission as a function of the amplitude of the trigger pulse (see Fig. 1(f)). The trigger pulse was tuned in resonance with the *XX-X* transition (see the red peak in Fig. 1(c)). The trigger pulse at the first minimum of the PL intensity corresponds to a $\pi$ pulse because the QD was at the upper energy level before the trigger pulse arrived. In addition to the pulse area, the delay time between the TPE and trigger pulses is another key parameter to achieve high excitation efficiency and indistinguishability. In theory, the trigger pulse should arrive immediately after the *XX* population generated by the TPE pulse reaches the maximum. In this case, the QD is completely deexcited from the *XX* state to the target *X* state by the trigger pulse and the time jitter induced by the spontaneous emission of *XX* is eliminated. In our experiment, the optimal delay time was found to be ~20 ps determined by directly measuring the *X* intensity as a function of the time interval between the TPE and trigger pulses and confirmed by the master equation simulation using QuTip[36] (see Supporting Fig. 4).

Next, we verified the ability of the DPE scheme to control the single-photon polarization. The experiment was divided into two parts. Firstly, we showed that the polarization of the *X* emission was determined by the trigger pulse. Secondly, we showed that the *X* emission was highly polarized. For the first part, we applied the DPE scheme to the QD and measured the vertically polarized component of the PL intensity at different linear polarization angles of the trigger pulse (see Fig. 2(a)). The "vertical" and "horizontal" polarization direction is defined by the neutral exciton eigenstates[37] (see the inset in Fig. 2(a)). The PL intensity reached the maximum (see the red arrow) when the trigger pulse was vertically polarized and dropped to the minimum (see the light blue arrow) when the trigger pulse was horizontally polarized. This result indicates that the polariza-tion of the QD emission was controlled by the trigger pulse. In between, the trigger pulse drove the QD to a superposition of the $X_H$ and $X_V$ states. The superposition state evolved with time and finally emitted a photon with the polarization determined by the time of the emission.

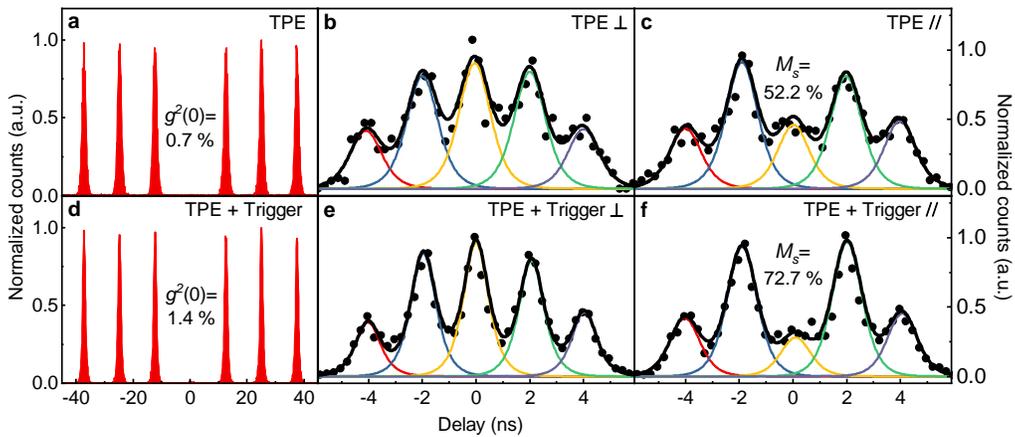

Figure 3. **a, d** HBT measurement of single-photon purity $(1 - g^2(0))$ performed under TPE (a) and DPE (d). **b, c, e, f** HOM measurement of interference visibility for photons emitted 2 ns apart. Normalized coincidence counts for cross- ($\perp$) and co-polarization (//) of the two interferometer arms, respectively, when the QD is excited by a TPE pulse (b), (c) and by TPE and trigger pulses (e), (f). Solid lines: fittings performed by convolving the exciton radiative lifetime with the instrument response function.

In order to show the QD emission under DPE was polarized, we fixed the polarization of the trigger pulse to vertical (horizontal) and measured the degree of linear polarization (DLP) by rotating a half waveplate (HWP) in front of a linear polarizer (see Fig. 2(b)). Under this condition, a DLP of 84(1)% (79(1)%) was measured. The imperfect DLP is caused by the birefringence of the optical components in the collection optical path and the slightly elliptically polarized cavity modes. To exclude these factors, we further characterized the QD emission using the Stokes vector (see Supporting Information Section 6) which gave a higher DOP of 89(1)% (86(1)%). The remaining imperfection of the DOP could be attributed to the incomplete population transfer from the XX state to the target X state by the trigger $\pi$ pulse caused mainly by the phonon-induced dephasing[38,39].

After characterizing the polarization of the QD emission, the next step is to examine the single-photon purity and indistinguishability. These parameters were evaluated by Hanbury Brown and Twiss (HBT) measurement and Hong–Ou–Mandel (HOM) measurement with a 2 ns delay time. In these experiments, the excitation lasers were blocked by combining spectral and polarization filtering. The trigger pulse and the polarizer in the collection optical path were set co-polarized with the $X_V$ state. The TPE pulse 0.45 meV detuned from the $X_V$ state was horizontally polarized. This configuration allowed the TPE pulse to be almost completely suppressed by the polarizer without blocking the QD emission. The trigger pulse 0.90 meV detuned from the $X_V$ state was then blocked by a grating-based spectral filter (see details in Methods). The full width at half maximum of the spectral filter was 0.07 meV, much larger than that of the X peak (0.01 meV). The single-photon purity $(1 - g^2(0))$ and raw indistinguishability $M_S$ measured under this configuration were 98.6(1)% and 73(5)%, respectively (see Figs. 3(d) – (f)). The corrected $M_S$ was 75(5)% taking into account the $g^2(0)$ and slightly unbalanced splitting ratio (48.8/51.2) of the fiber beam splitter in the HOM setup. In comparison, the single-photon purity and raw (corrected) indistinguishability measured with only a TPE pulse were 99.3(1)% and 52(5)% (53(5)%) (see Figs. 3(a) – (c)). The corrected $M_S$ under the DPE was improved by 21.8%, proving that the trigger pulse indeed recovered the reduced indistinguishability caused by spontaneous emission from the XX to X states[21]. The slight decrease of the single-photon purity under DPE was due to the residual trigger pulse leaking through the spectral filter. This fact implies that a large laser-QD detuning is highly desirable for completely blocking the laser using a spectral filter and hence obtaining near-unity single-photon purity.

The advantages of the DPE scheme can be further strengthened by an external magnetic field, namely, both the DOP of the QD emission and the laser-neutral exciton detuning can be significantly improved. The benefits of the magnetic field can be understood as follows. In the magnetic field, the two neutral exciton states ($X_H$ and $X_V$) initially with a small FSS split into two circularly polarized states $X_{\sigma+}$ and $X_{\sigma-}$ with a Zeeman splitting energy much larger than the linewidth of the QD emission (see Fig. 4(a)). This splitting allows to selectively collect the emission from only one purely circularly polarized exciton eigenstate, e.g. $X_{\sigma-}$. Additionally, the Zeeman splitting also helps to increase the energy difference between the G-XX transition and the $X_{\sigma-}$ transition, leading to an increased laser-neutral exciton detuning. In addition, the $X_{\sigma-}$ transition can be tuned in resonance with the cavity mode by the magnetic field, resulting in a shortened exciton lifetime and improved photon indistinguishability.

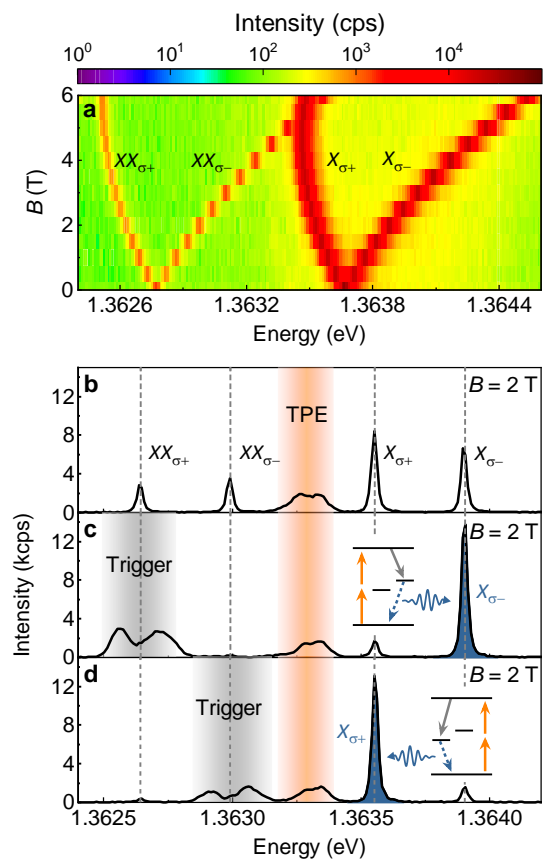

Figure 4. **a** Magnetic field-dependent PL mapping of the X and XX emission under above-barrier excitation. The magnetic field was applied parallelly to the QD growth axis (Faraday geometry). **b** PL spectrum of the QD excited by a TPE pulse at $B = 2$ T. **c** PL spectra under DPE with the trigger pulse resonant with the $XX_{\sigma+}$ peak. Inset: the XX cascade relaxation process. The QD was initially excited to the XX state by a TPE pulse. Then the trigger pulse drove the QD to the $X_{\sigma-}$ state. Finally, the QD emitted a left circularly polarized photon, corresponding to the $X_{\sigma-}$ peak at 1.3639 eV. **d** Following the same logic as (c), tuning the trigger pulse resonant with the $XX_{\sigma-}$ peak forced the QD to emit a right circularly polarized photon, corresponding to the $X_{\sigma+}$ peak at

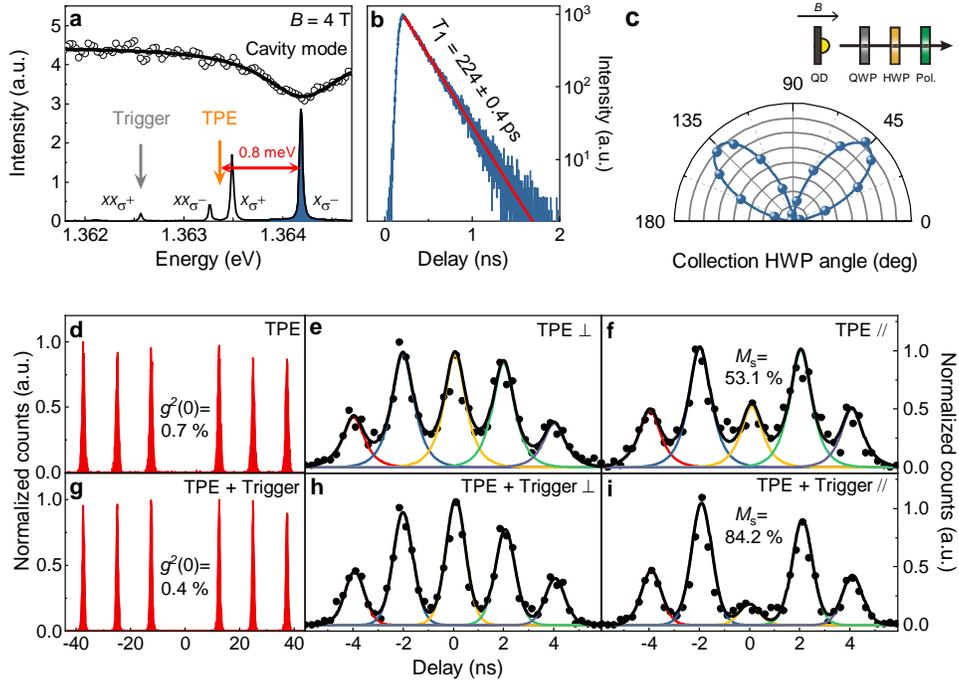

Figure 5. **a** The PL spectrum of the QD at $B = 4$ T under above-barrier excitation. The $X_{\sigma-}$ transition was tuned to the cavity resonance. The orange and gray arrows indicate the photon energy of TPE and trigger pulses that would be used in the following experiments performed with the DPE scheme. The TPE pulse was 0.81 meV detuned from the $X_{\sigma-}$ transition and the trigger pulse was resonant with the $XX_{\sigma+}$ peak. Open circles: the reflection spectrum of the cavity measured with a halogen lamp. Black line: fitting with a Lorentzian function. **b** PL decay of the $X_{\sigma-}$ emission under DPE. Red line: fitting with a single exponential function. **c** The intensity of the $X_{\sigma-}$ emission as a function of the HWP angle. Inset: the experimental setup for measuring the degree of circular polarization. **d, g** HBT measurement of the single-photon purity $(1 - g^2(0))$ for the $X_{\sigma-}$ emission performed under the TPE (d) and DPE (g) at $B = 4$ T. **e, f, h, i** HOM measurement of two-photon interference visibility for photons emitted 2 ns apart at $B = 4$ T. Normalized coincidence counts for cross- ($\perp$) and co-polarization ($//$) of the two interferometer arms, respectively, when the QD is excited by a TPE pulse (b), (c) and by TPE and trigger pulses (e), (f). Solid lines: fittings performed by convolving the exciton radiative lifetime with the instrument response function. QWP: quarter-wave plate.

1.3636 eV. Both the TPE and trigger pulses were linearly polarized. $\sigma+/\sigma-$: right/left circular polarization.

To demonstrate the magnetic-field-enhanced performance of the DPE scheme, we first studied the magneto-PL spectra of the QD. The magnetic field was applied along the growth axis of the QD (Faraday geometry). Fig. 4(a) shows the PL spectra measured with the above barrier excitation as a function of the magnetic field. Both the $X$ and $XX$ transitions split into two circularly polarized transitions. The shift of each transition was determined by the Zeeman splitting and the diamagnetic shift[37]. Fig. 4(b) shows the magneto-PL spectrum at $B = 2$ T measured under the two-photon excitation. Since all the $X$ and $XX$ transitions were spectrally separated, the selective excitation of the exciton eigenstates in the DPE scheme could be more clearly demonstrated. For example, when the trigger pulse was tuned in resonance with the $XX_{\sigma+}$ peak, the exciton emission was dominated by the $X_{\sigma-}$ peak (see Fig. 4(c)). Correspondingly, when the trigger pulse was tuned in resonance with the $XX_{\sigma-}$ peak, the exciton emission was dominated by the $X_{\sigma+}$ peak (see Fig. 4(d)). There was a small probability (~13%) that the exciton recombined via the unwanted neutral exciton states, shown as a minor peak in figs. 4(c) and (d). This observation can again be attributed to the incomplete population transfer from the $XX$ state to the target $X$ state.

The final step is to examine the single-photon purity and indistinguishability under the DPE in the magnetic field. To obtain the maximal indistinguishability with the current sample, the magnetic field was ramped to 4 T. At $B = 4$ T, the $X_{\sigma-}$ peak was resonant with the cavity (see Fig. 5(a)), shortening the $X_{\sigma-}$ lifetime by a factor of 1.8 to 224.0(4) ps (see Fig. 5(b)). Additionally, the detuning between the $X_{\sigma-}$ peak and the TPE pulse was increased to 0.81 meV (see Fig. 5(a)), resulting in a more efficient spectral filtering and in principle improved single-photon purity and indistinguishability. Furthermore, the magnetic field led to a purely circularly polarized exciton emission, confirmed by a DCP of the $X_{\sigma-}$ peak up to 99.6(7)% (see Fig.5 (c)). The measured DCP was limited by the birefringence of the optical elements in our setup and the slightly elliptical polarized cavity modes. Excluding these limiting factors by describing the polarization state using the Stokes vector yielded a DOP up to 101(2)% (see Supporting Information Section 6).

After we optimized the QD-cavity and laser-neutral exciton detunings, we moved to the HBT and HOM measurements. Both linearly polarized TPE and trigger pulses were blocked by a grating-based spectral filter and a linear polar-izer. Under this condition, a single-photon purity up to 99.6(1)% and raw (corrected) indistinguishability $M_S$ of 84(4)% (85(4)%) were obtained (see Figs. 5(g) - (i)). Compared with exciting the QD using only a TPE pulse, the corrected $M_S$ measured with the DPE was improved by 31.1% (see Figs. 5(d) – (f)). The single-photon purities measured under these two excitation conditions were quite close (<1%), limited by the dark counts of the single-photon detectors. The noise-limited ultralow $g^2(0)$ is a result of the two-photon excitation process that significantly supresses the re-excitation of the QD[40] and the ultrahigh laser extinction ratio ($10^{-8}$) by combined the spectral and polarization filtering (see detailed comparison in Supporting Information Section 7). The results above show that combining with the magnetic field, our new DPE scheme allows the generation of photons with relatively large laser-QD detuning, near-unity DOP, high single-photon purity and indistinguishability.

In the current work, although the excitation laser was spectrally separated from the neutral exciton transition with a detuning up to 0.81 meV, polarization filtering was still needed. This is because a spectral filter with a near-unity transmission rate, ultra-sharp edge, and sufficiently high laser extinction ratio is not readily available. This is not a problem for the measurement at 0 T, because the TPE pulse can be set to be cross-linearly polarized with the QD emission. Therefore, a polarizer in the collection optical path can almost completely suppress the TPE pulse without blocking the QD emission. However, the situation in the magnetic field is more complicated. In the magnetic field, the TPE pulse is linearly polarized, whereas the QD emission is circularly polarized. The linear polarizer needed to filter the TPE pulse will reduce the single-photon collection efficiency by 50%. This problem could be solved by increasing the laser-neutral exciton detuning using a QD with a biexciton binding energy $E_b$ of > 5 meV[32] or by applying a strong magnetic field. With sufficiently large laser-neutral exciton detuning, the excitation laser could be conveniently blocked solely using commercially available narrowband spectral filters with a transmission rate >90%. Alternatively, the strain or electric field could also be used to increase the biexciton binding energy[41–43] and split the two neutral exciton states into two well separated linearly polarized states[44–47]. In this case, single photon emission with a near-unity DOP could be efficiently collected even if a linear polarizer was needed.

In conclusion, we have demonstrated a DPE scheme to generate indistinguishable single photons with optically controlled polarization using spectrally separated excitation lasers. In this scheme, the TPE pulse detrimentally drives the QD to the biexciton state and the trigger pulse eliminates the time jitter caused by the relaxation from biexciton to the exciton state. With the help of the magnetic field, DOP up to 101(2)%, laser-neutral exciton detuning up to 0.81 meV, high single-photon purity (99.6(1)%) and indistinguishability of 85(4)% were achieved at $B$ = 4 T. Our work makes an important step towards a high-collection-efficiency on-demand indistinguishable SPSs, and shows that the magnetic field could be a powerful and versatile tool to boost the performance of the QD SPSs.

**Sample.** Our experiments were performed on an In(Ga)As/GaAs self-assembled QD embedded inside a 2-μm diameter micropillar cavity which has 26(18) λ/4-thick GaAs/Al$_{0.9}$Ga$_{0.1}$As mirror pairs forming the lower (upper) distributed Bragg reflectors (DBRs) (see the scanning electron microscope image in Supporting Fig. 2). The fabrication of the deterministically coupled QD-micropillar is based on the following steps. Firstly, the spatial and spectral information of the QDs are achieved by using the wide-field fluorescence imaging technique, then, high precision electron-beam lithography (EBL) is performed to define the resist pattern, and finally, the inductively-coupled plasma (ICP) reactive ion etching process is used to transfer the resist pattern of the pillar with certain diameter into the planar DBR sample.

**Optical setup.** The sample loaded in a closed-cycle cryostat (Attocube, AttoDry1000, $T$ = 3.7 K) was pumped and detected through a cage system-based confocal microscopy. For pulsed excitation, a Ti-sapphire laser (Coherent, Chameleon Ultra II) was used to generate 140-fs optical pulses every 12.5 ns, and then the femtosecond pulses were sent into two pulse shapers to pick out TPE (10 ps duration) and trigger (15 ps duration) pulse. The interval between the TPE pulse and the trigger pulse was 20 ps (15 ps) for 0 T (4 T) measurement and introduced by an optical delay line. For lifetime measurement, the single photons were detected using a superconducting nanowire single-photon detector (timing jitter: ~60 ps, see the instrument response function (IRF) in Supporting Fig. 3a) and a time-correlated single-photon counting module (PicoQuant, HydraHarp400). For second-order correlation measurements, the QD emission passed through a grating-based filter (bandwidth: ~0.07 meV, transmission rate: 39%) and then was sent to an HBT-type setup or 2 ns delay Mach–Zehnder interferometer and finally detected by two single-photon avalanche diodes (Excelitas, SPCM-AQRH-15, timing jitter: ~427 ps, see the IRF in Supporting Fig. 3b) (for further details see Supporting Information Section 5).

## ASSOCIATED CONTENT

### Supporting Information

The Supporting Information is available free of charge on the ACS Publications website.
Fine-structure splitting of the quantum dot, characterization of the micropillar cavity, time resolution of single-photon detectors, effect of the trigger pulse, experimental setup, Stokes polarization parameters, extinction ratio of spectral and polarization filter, collection efficiency (PDF).

## AUTHOR INFORMATION

### Corresponding Author

**Feng Liu** − College of Information Science and Electronic Engineering, Zhejiang University, Hangzhou 310027, China;
Email: feng_liu@zju.edu.cn;
**Ying Yu** − School of Electronics and Information Technology, Sun Yat-sen University, Guangzhou 510275, China;

Email: yuying26@mail.sysu.edu.cn.

**Author Contributions**

‡(J.Y., S.L.) These authors contributed equally.

**Notes**

The authors declare no competing financial interests.

## ACKNOWLEDGMENT

This work was supported by the National Natural Science Foundation of China (62075194, 11934011, 61574138, 61974131, 12074442), Natural Science Foundation of Zhejiang Province (LGJ21F050001), Major Scientific Research Project of Zhejiang Lab (2019MB0AD01), Fundamental Research Funds for the Central Universities (2021QNA5006).

TOC Graphic:

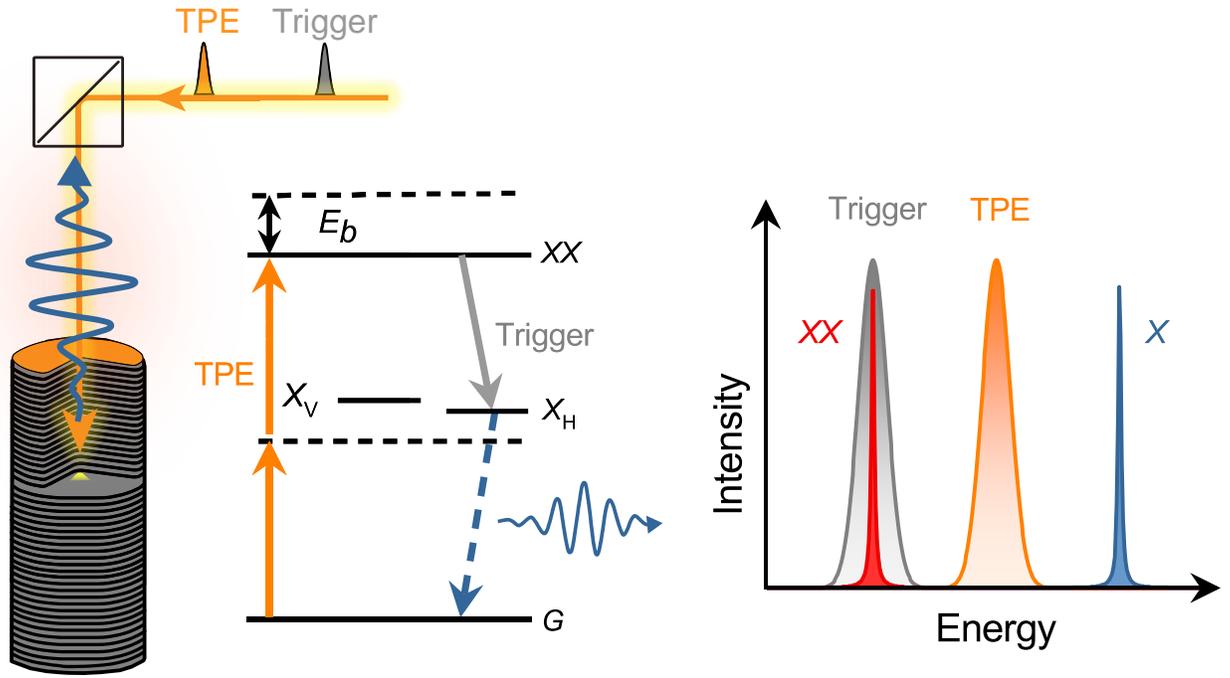

# Supplementary Information for

# "Double-pulse generation of indistinguishable single photons with optically controlled polarization"

Junyong Yan[1,‡], Shunfa Liu[2,‡], Xing Lin[1], Yongzheng Ye[1,3], Jiawang Yu[1,4], Lingfang Wang[1], Ying Yu[2,*], Yanhui Zhao[2,5], Yun Meng[6,7], Xiaolong Hu[6,7], Dawei Wang[1,8], Chaoyuan Jin[1,4], Feng Liu[1,4,*]

[1]Interdisciplinary Center for Quantum Information, State Key Laboratory of Modern Optical Instrumentation, College of Information Science and Electronic Engineering, Zhejiang University, Hangzhou 310027, China.

[2]State Key Laboratory of Optoelectronic Materials and Technologies, School of Physics, School of Electronics and Information Technology, Sun Yat-sen University, Guangzhou 510275, China

[3]College of Optical Science and Engineering, Zhejiang University, Hangzhou 310027, China.

[4]International Joint Innovation Center, Zhejiang University, Haining 314400, China

[5]School of Physics and Optoelectronic Engineering, Ludong University, Yantai, Shandong 264025, China.

[6]School of Precision Instrument and Optoelectronic Engineering, Tianjin University, Tianjin 300072, China.

[7]Key Laboratory of Optoelectronic Information Science and Technology, Ministry of Education, Tianjin 300072, China.

[8]Zhejiang Province Key Laboratory of Quantum Technology and Device, Department of Physics, Zhejiang University, Hangzhou 310027, China.

## 1. Fine-structure splitting of the quantum dot

To obtain the fine-structure splitting (FSS) energy of the neutral exciton states, a series of QD PL spectra were acquired at different linear polarization angles as shown in Supplementary Fig. 1a. By fitting PL spectra with a gaussian function, we obtained the polarization dependence of the neutral exciton PL photon energy (see Supplementary Fig. 1b). Fitting the photon energy with a sinusoidal function gave an exciton FSS of 5.9(1) μeV.

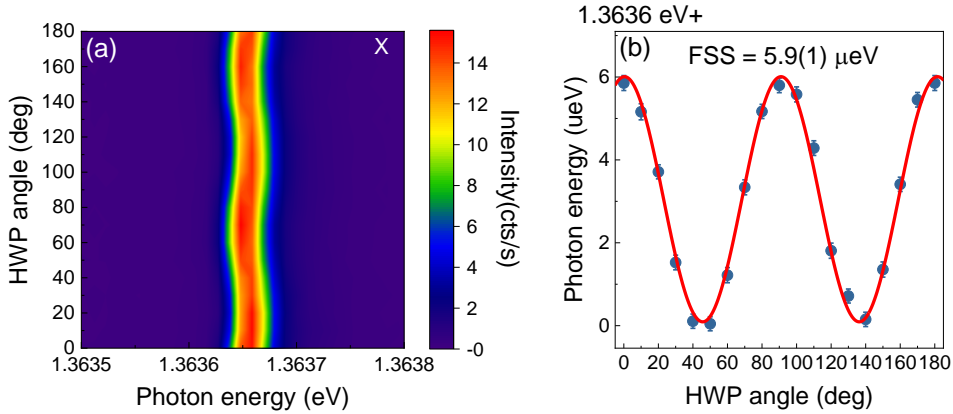

**Supplementary Figure 1 | FSS of the QD. (a)** PL mapping of the neutral exciton states as a function of the angle of HWP in front of a liner polarizer in the collection optical path. **(b)** The polarization dependence of the neutral exciton PL photon energy. Redline: fitting with a sinusoidal function.

## 2. Characterization of the micropillar cavity

Our experiments were performed on an In(Ga)As/GaAs QD embedded inside a 2-μm-diameter micropillar cavity which contained 26(18) λ/4-thick GaAs/Al$_{0.9}$Ga$_{0.1}$As mirror pairs (as shown in Supplementary Fig. 2a.). To characterize the cavity mode, we illuminated the sample with a halogen lamp and measured reflection spectra. A dip in the reflection spectra corresponds to the cavity mode (Supplementary Fig. 2b.). Fitting the spectrum with a Lorentzian function gave a Q factor of 1351. In theory, a micropillar cavity with cylindrical symmetry has two degenerate circularly polarized fundamental modes. To investigate the energy splitting and polarization of the cavity modes, we measured the reflection spectra as a function of the detected polarization angle. Supplementary Fig. 2c. shows the peak energy and area of the reflection spectra fitted with a Lorentzian function. The nearly constant peak energy indicates that the two fundamental modes are highly degenerate. The slightly varying peak area indicates that the fundamental modes are slightly elliptically polarized.

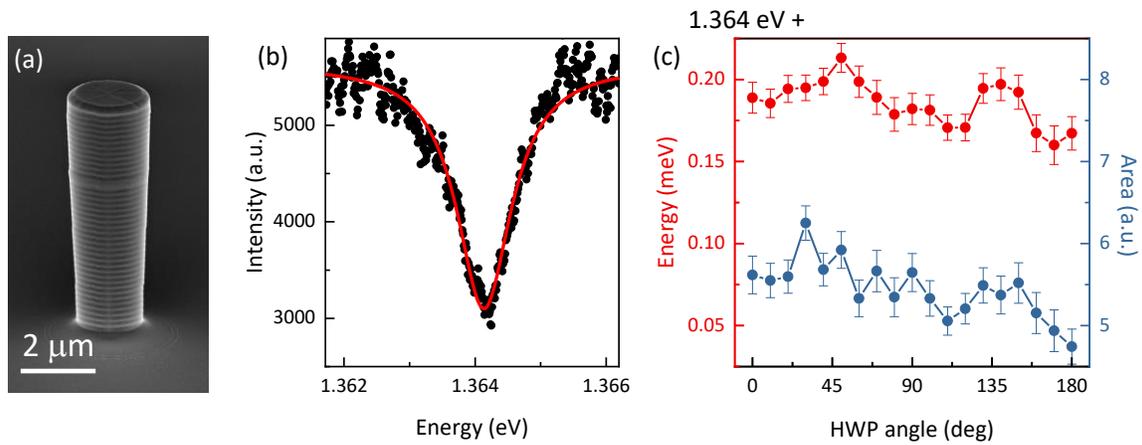

**Supplementary Figure 2 | Characterization of the micropillar cavity.** (**a**) Scanning electron microscope (SEM) image of a 2-µm-diameter micropillar cavity. (**b**) The reflection spectrum of the cavity (Q = 1351) measured with a halogen lamp. (**c**) Polarization-resolved measurement of the fundamental cavity modes.

### 3. Time resolution of single-photon detectors

Two types of single-photon detectors were used in our experiments, a superconducting nanowire photon detector (SNSPD) was used in time-resolved PL measurements, and two single-photon avalanche diodes (SPAD) were used for HBT and HOM measurements. Instrument response functions of two experimental setups measured with a femtosecond pulsed laser are shown in Supplementary Fig. 3. The full width at half-maximum of the IRFs were $60.0(1)$ ps and $427(2)$ ps for detection systems with the SNSPD and SPAD, respectively.

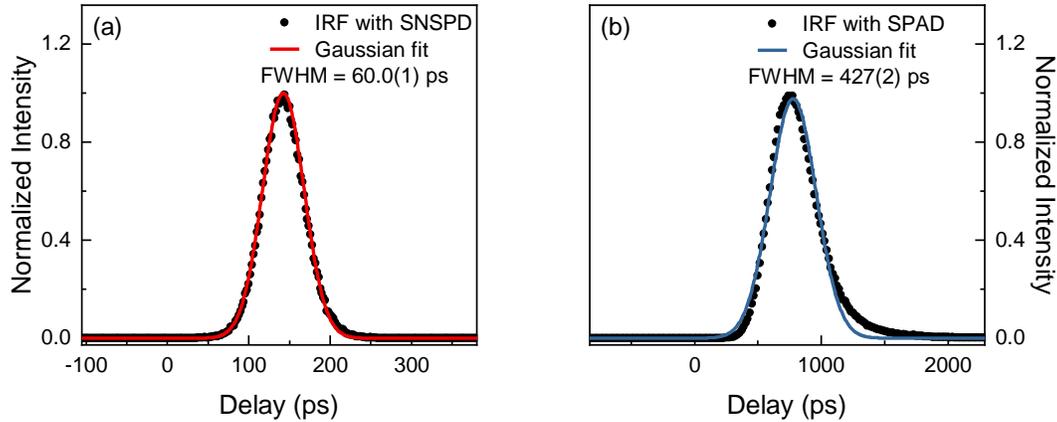

**Supplementary Figure 3 | The instrument response function (IRF) of the detection system with (a) SNSPD and (b) SPAD.**

### 4. Effect of the trigger pulse

As already mentioned in the main text, a trigger pulse resonant with $X$-$XX$ transition enables deterministic preparation of $X$ eigenstates with negligible time jitter. Setting the trigger pulse to be parallel to the detected polarization (V, in this experiment), we demonstrated that the trigger pulse could deterministically prepare an exciton eigenstate to double the photon collection efficiency as shown in the PL map in Supplementary Fig. 4. At fixed pulse area, with the increase of the pulse interval, the collected $X$ intensity decayed exponentially because the $XX$ relaxed to one of the two $X$ eigenstates randomly via spontaneous emission (see the upper panel in Supplementary Fig. 4). At fixed TPE-trigger interval, as the pulse area of the trigger pulse increased, the $X$ intensity oscillating with the trigger pulse amplitude, reflecting the Rabi oscillation of the $X_V$-$XX$ transition (see the right panel in Supplementary Fig. 4).

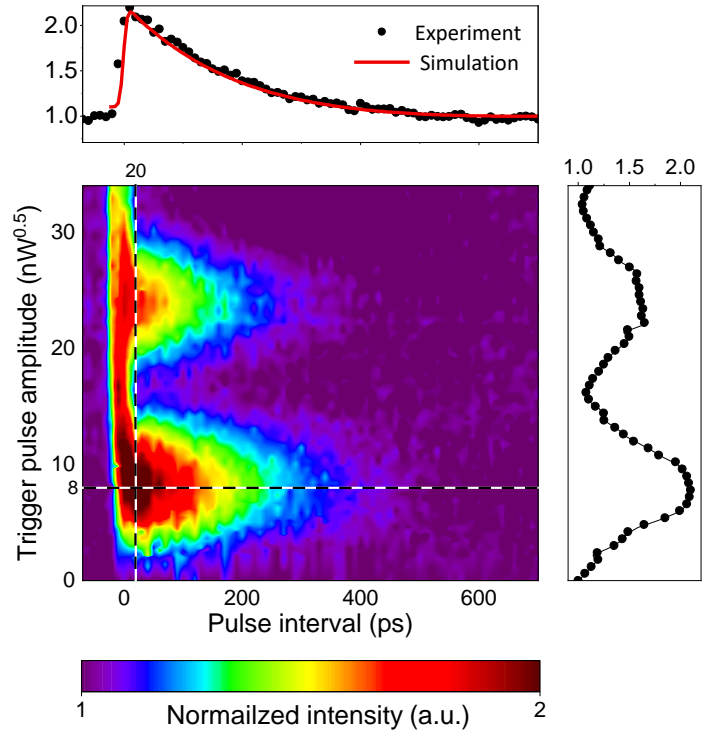

**Supplementary Figure 4 | The intensity of *X* emission versus the trigger pulse area and TPE-trigger interval at *B* = 0 T.** The pulse area of the TPE pulse was π. The trigger pulse polarization was set to co-polarized with the detected polarization. Upper panel: the intensity of the *X* emission versus the TPE-trigger interval at fixed trigger pulse power of 64 nW. Red curve: the simulated intensity of QD emission using a Lindblad master equation which was solved and analyzed with the help of the Quantum Toolbox in Python (QuTiP). Right panel: the intensity of the *X* emission versus area of the trigger pulse with a TPE-trigger interval of 20 ps.

## 5. Experimental setup

The experimental setup used in this work is shown in Supplementary Fig. 5. The optical experiments were performed using a confocal microscope with the sample loaded in a closed-cycle cryostat and excited by two picosecond pulses with variable time delay. The QD emission was collected by a single-mode fiber and sent to either a spectrometer for the measurement of PL spectra or an HBT/HOM setup for the measurement of single-photon purity and indistinguishability.

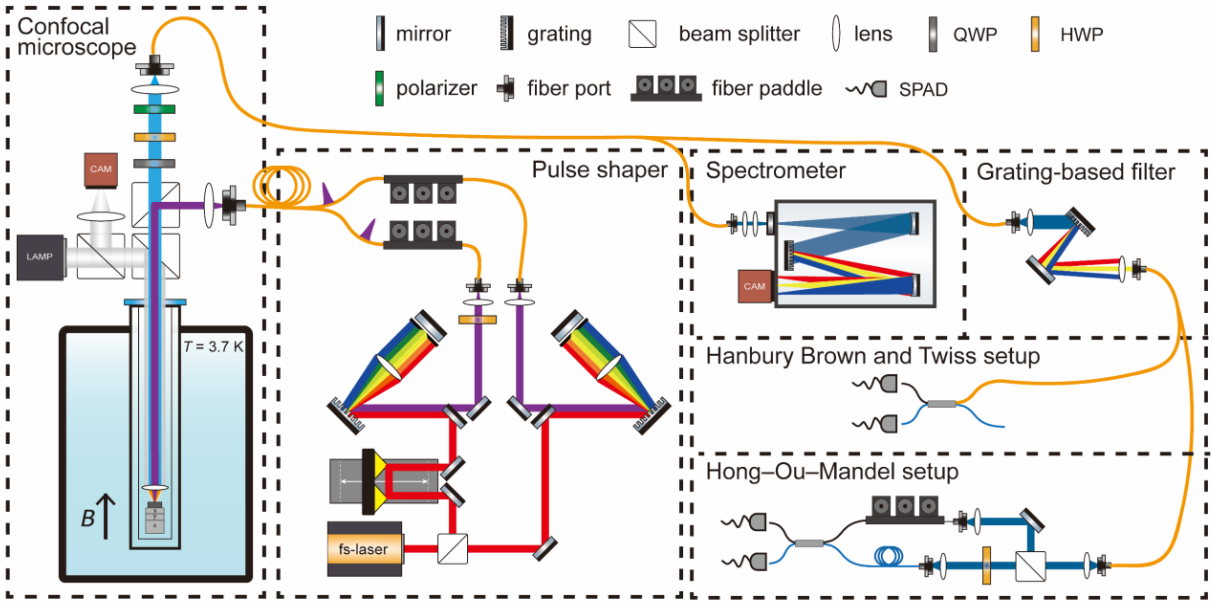

**Supplementary Figure 5 | Schematic of the setup for optical characterizations.** Left panel: a confocal micro-PL setup with the sample loaded in a closed-cycle cryostat ($T$ = 3.7 K). Central panel: two pulse shapers based on folded 4f systems, optical delay line, and laser polarization control. Right panel: Spectrometer for measuring PL spectra. Grating-based filter (full width at half-maximum: ~0.07 meV) for blocking the DPE pulses. HBT (HOM) setup for single-photon purity (indistinguishability) measurement. QWP: quart-wave plate, HWP: half-wave plate.

## 6. Stokes polarization parameters

To describe the single-photon polarization more accurately, we measured the Stokes polarization parameters using a rotating QWP method. The experiment setup is shown in Supplementary Fig. 6. Firstly, we fixed the axis of the polarizer and the fast axis of the QWP to the horizontal direction. Secondly, we rotated the QWP by 180° in a step of 5° and recorded the intensity of the QD emission as a function of the QWP angle. Thirdly, we used the Fourier coefficient equation (as shown in Equation (1)) to get the parameters A, B, C and D. Finally, we used Equation (2) to get the normalized Stokes parameters and degree of polarization (DOP) as shown in Supplementary Table 1.

**Supplementary Table 1. Stokes parameters and DOP.**

| Parameters | $B = 0$ T, Trigger H | $B = 0$ T, Trigger V | $B = 4$ T, Trigger $X_{\sigma-}$ |
|---|---|---|---|
| $S_0$ | 1.00 | 1.00 | 1.00 |
| $S_1$ | 0.79(3) | -0.82(1) | 0.06(3) |
| $S_2$ | -0.28(5) | 0.30(6) | 0.24(1) |
| $S_3$ | -0.16(1) | 0.16(1) | -0.98(2) |
| DOP | 0.86(1) | 0.89(1) | 1.01(2) |

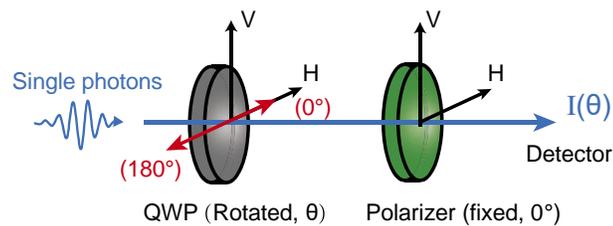

**Supplementary Figure 6 | Schematic of rotating QWP method.**

$$\begin{cases} A = \dfrac{2}{N}\sum_{n=1}^{N} I(\theta_n) \\ B = \dfrac{4}{N}\sum_{n=1}^{N} I(\theta_n)\sin(2\theta_n) \\ C = \dfrac{4}{N}\sum_{n=1}^{N} I(\theta_n)\cos(4\theta_n) \\ B = \dfrac{4}{N}\sum_{n=1}^{N} I(\theta_n)\sin(4\theta_n) \end{cases} \quad (1)$$

$$\begin{cases} S_0 = 1 \\ S_1 = \dfrac{2C}{A-C} \\ S_2 = \dfrac{2D}{A-C} \\ S_3 = \dfrac{B}{A-C} \\ DOP = \dfrac{\sqrt{S_1^2+S_2^2+S_3^2}}{S_0} \end{cases} \quad (2)$$

### 7. The extinction ratio of spectral and polarization filter

One of the important advantages of the DPE scheme is its compatibility with both polarization and frequency filtering. The spectral extinction ratio of TPE (Trigger) pulse is ~$10^{-4}$ ($10^{-6}$) in $B = 0$ T configuration, and ~$10^{-5}$ ($10^{-6}$) in $B = 4$ T configuration (as shown in Supplementary Fig. 7 (b)). The extinction ratio of the cross-polarization setup is ~$10^{-4}$ (as shown in Supplementary Fig. 7 (c)). Combined with polarization and frequency filtering, the extinction ratio of TPE pulse can reach ~$10^{-8}$.

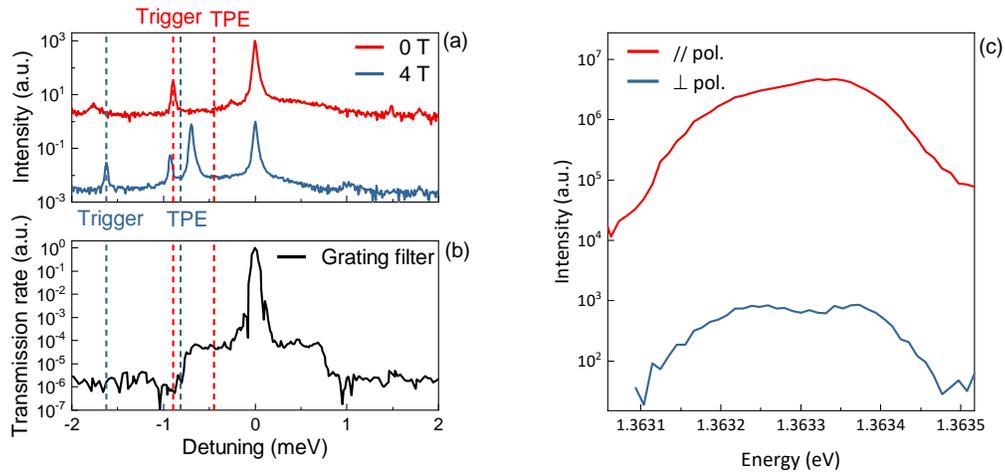

**Supplementary Figure 7 | Extinction ratio of spectral and polarization filter.** (a) PL spectra at $B = 0$ and 4 T under above-barrier excitation. (b) The normalized transmission of homemade grating-based filter measured by sweeping a narrow band continuous-wave laser. (c) The TPE $\pi$ pulse scattering intensity when the microscope was set at parallel and orthogonal configuration.

## 8. Collection efficiency

The coupling efficiency/transmission rate of all the optical elements in our setup are listed in Supplementary Table 2.

The collection efficiency defined as the ratio of the QD emission coupled into a single-mode fiber is 0.63% in our experiment. The collection efficiency with conventional resonance fluorescence (RF) technique where the exciton is resonantly excited and the laser is filtered out by the polarizer is 0.81%. Increasing the biexciton binding energy of the QD would allow us to replace the grating-based spectral filter (transmission rate: 39%) with a commercially available ultra-narrow bandpass spectral filter (Alluxa) with a transmission rate above 90%, leading to an expected collection efficiency of 1.46%.

**Supplementary Table 2. The coupling efficiency/transmission rate of optical elements in the setup.**

| Optical elements | Coupling efficiency/transmission rate | |
|:---:|:---:|:---:|
| | Conventional RF | DPE |
| Sample to objective | 9.0% | 9.0% |
| Objective | 88.9% | 88.9% |
| Window | 92.6% | 92.6% |
| 50:50 BS | 49.8% | 49.8% |
| HWP | 98.4% | 98.4% |
| Polarizer | 40.7% | 81.4% |
| Fiber coupling | 55.0% | 55.0% |
| Spectral filter | N/A | 39.0% (90.0% [a]) |
| **Overall collection efficiency** | **0.81%** | **0.63% (1.46% [a])** |

[a] The expected collection efficiency after replacing the grating-based filter with a commercially available ultra-narrow bandpass spectral filter.